\documentclass[preprint,aps,12pt,showpacs,nofootinbib,tightenlines]{revtex4}
\usepackage{amsmath}
\usepackage{amssymb}
\usepackage{epsfig}
\usepackage{graphicx}
\begin{document}
%%%%%%%%%%%%%%%%%%%%%%%%%%%%%%%%%%%%%%%%%%%%%
%%%%%%%%%%%%%%%%%%%%%%%%%%%%%%%%%%%%%%%%%%%%%%%%%%%%%%
\def\pslash{\rlap{\hspace{0.02cm}/}{p}}
\def\eslash{\rlap{\hspace{0.02cm}/}{e}}
%%%%%%%%%%%%%%%%%%%%%%%%%%%%%%%%%%%%%%%%%%%%%%%%%%%%%%
\title{The associated productions of the new gauge boson $B_{H}$ in the littlest Higgs model with a
SM gauge boson via $e^+e^-$ collision}
\author{Xuelei Wang} \email{wangxuelei@sina.com}
\author{Qingguo Zeng}
\author{Zhenlan Jin}
\author{Suzhen Liu}
\affiliation{College of Physics and Information Engineering, Henan
Normal University, Xinxiang, Henan 453007. P.R. China}
\thanks{This work is supported by the National Natural Science
Foundation of China(Grant No.10375017 and No.10575029).}

\date{\today}
\vspace{3cm}
\begin{abstract}
\indent With the high energy and luminosity, the planned ILC has
the considerable capability to probe the new heavy particles
predicted by the new physics models. In this paper, we study the
potential to discover the lightest new gauge boson $B_{H}$ of the
littlest Higgs model via the processes $e^+e^- \rightarrow \gamma
(Z)B_H$ at the ILC. The results show that the production rates of
these two processes are large enough to detect $B_H$ in a wide
range of the parameter space, specially for the process $e^+e^-
\rightarrow \gamma B_H$. Furthermore, there exist some decay modes
for $B_H$ which can provide the typical signal and clean
background. Therefore, the new gauge boson $B_H$ should be
observable via these production processes with the running of the
ILC if it exists indeed.
\end{abstract}


\maketitle
\newpage
\section{Introduction}
\hspace{0.6cm}

A simple double scalar field yields a perfectly appropriate gauge
symmetry breaking pattern in the standard model(SM). However, one
well-known difficulty is that the mass of the Higgs boson receives
quadratic loop corrections and such corrections become large at a
high energy scale which is known as hierarchy problem. In order to
achieve an effective Higgs boson mass on the order of 100 GeV, as
required by fits to precision electroweak
parameters\cite{precision}, new physics at the TeV scale is
therefore needed to cancel the quadratic corrections in the SM.
The possible new physics scenarios at the TeV scale might be
supersymmetry(SUSY)\cite{SUSY}, extra dimension\cite{extra}, and
the technicolor(TC) model\cite{TC} etc. Recently, there has been a
new formulation for the physics of electroweak symmetry breaking,
dubbed the "little Higgs" models\cite{little,LH}, which offer a
very promising solution to the hierarchy problem in which the
Higgs boson is naturally light as a result of nonlinearly realized
symmetry. The key idea of the little Higgs theory may be that the
Higgs boson is a Goldstone boson which acquires mass and becomes
the pseudo-Goldstone boson via symmetry breaking at the
electroweak scale and remains light, being protected by the
approximate global symmetry and free from 1-loop quadratic
sensitivity to the cutoff scale. Such models can be regarded as
the important candidates of new physics beyond the SM. The
littlest Higgs(LH) model\cite{LH}, based on a $SU(5)/SO(5)$
nonlinear sigma model, is the simplest and phenomenologically
viable model to realize the little Higgs idea. It consists of a
$SU(5)$ global symmetry, which is spontaneously broken down to
$SO(5)$ by a vacuum condensate $f$. At the same time, the gauge
subgroup $[SU(2)\times U(1)]^{2}$ is broken to its diagonal
subgroup $SU(2)\times U(1)$, identified as the SM electroweak
gauge group. In such breaking scenario, four new massive gauge
bosons( $B_{H},~Z_{H},~W^{\pm}_{H}$) are introduced and their
masses are in the range of a few TeV, except for $B_{H}$ in the
range of hundreds of GeV. The existence of these new particles
might provide the characteristic signatures at the present and
future high energy collider experiments\cite{Han, signatures} and
the observation of them can be regarded as the reliable evidence
of the LH model.

\indent On the experimental aspect, although the hadron colliders
Tevatron and LHC can play an important role in probing the new
particles predicted by the new physics models, the search for new
particles has strongly motivated projects at future high energy
$e^+e^-$ linear collider, i.e., the International Linear Collider
(ILC), with the center of mass(c.m) energy $\sqrt{s}$ =300 GeV-1.5
TeV and the integrated luminosity 500 $fb^{-1}$ within the first
four years\cite{ILC}. With high luminosity and clean environment,
the most precise measurements will be performed at the ILC.

The ILC will provide us a good chance to probe the new gauge
bosons in the LH model and some production processes of these new
gauge bosons at the ILC have been studied\cite{production, yue,
wang}. As we know, with the mass of hundreds GeV, the gauge boson
$B_H$ is the lightest new particle in the LH model and it is light
enough to be produced at the first running of the ILC. So the
exploration of $B_{H}$ at the ILC would play an important role in
testing the LH model. We have studied some $B_{H}$ production
processes at the photon collider, i.e., $\gamma\gamma\rightarrow
W^{+}W^{-}B_H$ and $e^{-}\gamma\rightarrow \gamma(Z)e^{-} B_{H}$
\cite{wang}. As we know, the photon collider has advantages in
probing the new particles in the new physics models. Our studies
show that the sufficient typical $B_H$ events could be detected at
the photon collider. However, with the running of the ILC, $B_H$
might be first discovered via $e^+e^-$ collision if it exists
indeed, and the study of the $B_H$ productions via $e^+e^-$
collision is more imperative. In this paper, we study two
interesting $B_H$ production processes via $e^+e^-$ collision,
i.e., $e^+e^- \rightarrow \gamma B_H$ and $e^+e^- \rightarrow
ZB_H$ at the ILC. These processes are particularly interesting in
various aspects. From an experimental point of view, these
processes can produce enough $B_H$ signals with clean background
and the final states are easy to detect. Furthermore, these
processes can be realized at the first running of the ILC. From
the theoretical point of view, these processes have a simple
structure providing clean tests of the properties of the
$B_He^+e^-$ coupling.

 \indent  The rest parts of this paper are organized as
follows. In section II, we first present a brief review of the LH
model and then give the production amplitudes of the processes.
the numerical results and conclusions are given in Section III.

\section{The processes of $e^+e^- \rightarrow \gamma (Z) B_H$}
\subsection{ A brief review
of the LH model}

The LH model is one of the simplest and phenomenologically viable
models to realize the little Higgs idea. The LH model is embedded
into a non-linear $\sigma-$model with the corset space of
$SU(5)/SO(5)$. At the scale $\Lambda_{s}\sim{4\pi}f$, the vacuum
condensate scale parameter $f$ breaks the global $SU(5)$ symmetry
into its subgroup $SO(5)$ resulting in 14 Goldstone bosons. The
effective field theory of these Goldstone bosons is parameterized
by a non-linear $\sigma$-model with a gauge symmetry $[SU(2)\times
U(1)]^{2}$, and the $[SU(2)\times U(1)]^{2}$ gauge symmetry is
broken to the diagonal $SU(2)_{L}\times U(1)_{Y}$ subgroup which
is identified as the electroweak gauge symmetry. The effective
non-linear lagrangian which is invariant under the local gauge
group $[SU(2)\times U(1)]^{2}$ can be written as
\begin{eqnarray}
 \mathcal{L}_{eff}=\mathcal{L}_{G}+\mathcal{L}_{F}+\mathcal{L}_{\Sigma}
 +\mathcal{L}_{Y}-V_{CW}(\Sigma).
\end{eqnarray}
 Where $\mathcal{L}_{G}$ consists of the pure gauge terms;
  $\mathcal{L}_{F}$ is the fermion kinetic terms,
  $\mathcal{L}_{\Sigma}$ consists of the $\sigma$-model terms of the LH
model, $\mathcal{L}_{Y}$ is the Yukawa couplings of fermions and
pseudo-Goldstone bosons, and $V_{CW}(\Sigma)$ is the
Coleman-Weinberg potential generated radiatively
 from $\mathcal{L}_{Y}$ and $\mathcal{L}_{\Sigma}$.
   The scalar fields are parameterized by
\begin{eqnarray}
 \Sigma(x)=e^{2i\Pi/f}\Sigma_0,
\end{eqnarray}
with $\langle\Sigma_{0}\rangle\sim $ f, which generates the masses
and mixing between the gauge bosons. The leading order
dimension-two term for the scalar sector in the non-linear
$\sigma$-model can be written as
\begin{eqnarray}
{\pounds}_{\Sigma}=\frac{f^2}{8}Tr\{(D_{\mu}\Sigma)(D^{\mu}\Sigma)^{+}\},
\end{eqnarray}
with the covariant derivative of $\Sigma$ given by
\begin{eqnarray}
 D_{\mu}\Sigma=\partial_{\mu}\Sigma-i\sum^{2}_{j=1}[g_j(W_{\mu j}\Sigma+\Sigma W^T_{\mu j})+
 g'_j(B_{\mu j}\Sigma+\Sigma B^T_{\mu j})].\nonumber
\end{eqnarray}
Where $g_{j}$ and $g'_{j}$ are the couplings of the $[SU(2)\times
U(1)]$ groups, respectively. $W_{\mu j}=\sum^3_{a=1}W^a_{\mu
j}Q^a_j$
 and $B_{\mu j}=B_{\mu j}Y_j$  are the $SU(2)$ and $U(1)$ gauge fields,
respectively.

  The spontaneous gauge symmetry breaking thereby gives the gauge
boson mass eigenstates
\begin{eqnarray}
W_{\mu}=sW_{\mu 1}+cW_{\mu 2}, ~~~~~~W'_{\mu}=-cW_{\mu 1}+sW_{\mu
2},
\\ \nonumber
B_{\mu}=s'B_{\mu 1}+c'B_{\mu 2},~~~~~~B'_{\mu}=-c'B_{\mu
1}+s'B_{\mu 2}.
\end{eqnarray}
The gauge bosons W and B are massless states identified as the SM
gauge bosons, with couplings $g=g_{1}s=g_{2}c$ and
$g'=g_{1}'s'=g_{2}'c'$. In this paper, the vacuum condensate scale
parameter $f$, the mixing parameters $c'$ and $c$ between the
charged and neutral vector bosons are the free parameters.

 Through radiative corrections, the gauge, the Yukawa, and
 self-interactions of the Higgs field generate a Higgs potential
 which triggers the EWSB. Now the SM gauge bosons $W$ and $Z$ acquire masses
 of order $v$, and small (of order $v^2/f^2$) mixing between the heavy
 gauge bosons and the SM gauge bosons $W,~Z$ occurs. The masses of
 $W,~Z$ and their couplings to the SM particles
 are modified from those in the SM at the order of $v^2/f^2$.
  In the
 following, we denote the mass eigenstates of SM gauge fields by
 $W_{L}^{\pm}, Z_{L}$ and  $A_{L}$ and the new heavy gauge bosons by $W_{H}^{\pm}, Z_{H}$ and
$B_{H}$. The masses of the neutral gauge bosons are given to
$\mathcal {O}(\nu^{2}/f^{2})$  by \cite{Han, production}

\begin{eqnarray}
M^{2}_{A_{L}}&=&0, \\ \nonumber
M^{2}_{B_{H}}&=&(M^{SM}_{Z})^{2}s^{2}_{W}\{\frac{f^{2}}{5s'^2c'^2v^2}-1
+\frac{v^2}{2f^2}[\frac{5(c'^2-s'^2)^2}{2s^2_W}-\chi_H\frac{g}{g'}\frac{c'^2s^2+c^2s'^2}{cc'ss'}]\},
\\ \nonumber
M^{2}_{Z_{L}}&=&(M^{SM}_{Z})^{2}\{1-\frac{v^{2}}{f^{2}}[\frac{1}{6}+\frac{1}{4}(c^{2}-s^{2})^{2}+
\frac{5}{4}(c'^{2}-s'^{2})^{2}]+8\frac{v'^2}{v^2}\},
\\ \nonumber
M^{2}_{Z_{H}}&=(&M^{SM}_{W})^{2}\{\frac{f^2}{s^2c^2v^2}-1+\frac{v^2}{2f^2}[\frac{(c^2-s^2)^2}{2c_W^2}
+\chi_H\frac{g'}{g}\frac{c'^2s^2+c^2s'^2}{cc'ss'}]\},
\end{eqnarray}
with
 $\chi_{H}=\frac{5}{2}gg'\frac{scs'c'(c^{2}s'^{2}+s^{2}c'^{2})}{5g^{2}s'^{2}c'^{2}-g'^2s^{2}c^{2}}$.
Where $v$=246 GeV is the elecroweak scale, $v'$ is the vacuum
expectation value of the scalar $SU(2)_{L}$ triplet,
$c_{W}=\cos\theta_{W}$ and $s_{W}=\sin\theta_{W}$ represent the
weak mixing angle.

 In the LH model, the relevant
couplings of the neutral gauge bosons to the electron pair can be
written in the form
$\Lambda_{\mu}^{V_{i}ee}=i\gamma_{\mu}(g_{V}+g_{A}\gamma^{5})$
with\cite{Han,Buras}
\begin{eqnarray}
g_{V}^{B_{H}ee}&=&\frac{g'}{2s'c'}(2y_e-\frac{9}{5}+\frac{3}{2}c'^{2}),\\
\nonumber
      g_{A}^{B_{H}ee}&=&\frac{g'}{2s'c'}(-\frac{1}{5}+\frac{1}{2}c'^{2}),\\
      \nonumber
g_{V}^{Z_{L}ee}&=&-\frac{g}{2c_{W}}\{(-\frac{1}{2}+2s^{2}_{W})-\frac{v^2}{f^2}[-c_w\chi_Z^{W'}c/2s
+\frac{s_w\chi_Z^{B'}}{s'c'}(2y_e-\frac{9}{5}+\frac{3}{2}c'^2)]\},\\
\nonumber
g_{A}^{Z_{L}ee}&=&-\frac{g}{2c_{W}}\{\frac{1}{2}-\frac{v^2}{f^2}[c_W\chi_Z^{W'}c/2s
+\frac{s_W\chi_Z^{B'}}{s'c'}(-\frac{1}{5}+\frac{1}{2}c'^2)]\},
\\ \nonumber
g_{V}^{\gamma ee}&=&-e, ~~~~~~~~~~~g_{A}^{\gamma ee}=0.\nonumber
\end{eqnarray}
Where, $\chi_Z^{B'}=\frac{5}{2s_W}s'c'(c'^2-s'^2)$ and
$\chi_Z^{W'}=\frac{1}{2c_W}sc(c^2-s^2)$. The U(1) hypercharge of
electron, $y_e$, can be fixed by requiring that the $U(1)$
hypercharge assignments be anomaly free, i.e.,
   $y_e=\frac{3}{5}$. This is only one example among several alternatives for
the U(1) hypercharge choice\cite{Han,Csaki}.

\subsection{The production amplitudes of the processes $e^+e^- \rightarrow \gamma (Z)B_H$}
As we have mentioned above, the lightest $B_H$ should be the first
signal of the LH model. With the coupling $e^+e^-B_H$, $B_H$ can
be produced associated with a neutral SM gauge boson $\gamma$ or
$Z$ at tree-level via $e^+e^-$ collision, i.e., $e^+e^-\rightarrow
\gamma (Z)B_H$. The relevant Feynman diagrams of the processes are
shown in Fig.1.
\begin{figure}
\begin{center}
\includegraphics [scale=0.7] {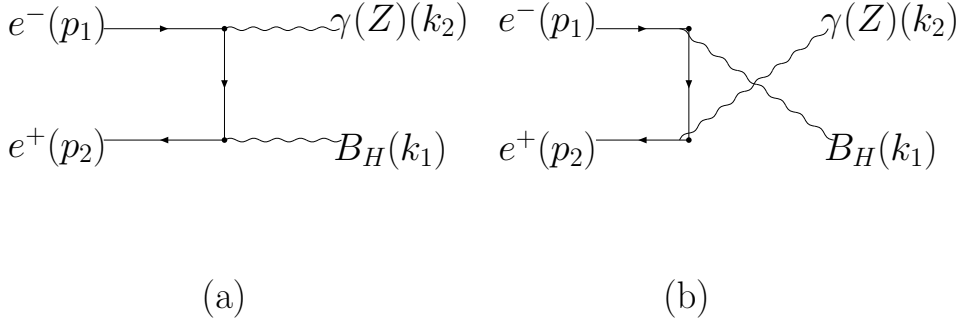}
\vspace{-14cm}
 \caption{The Feynman diagrams of the processes
$e^+e^-\rightarrow \gamma (Z)B_H$.} \label{fig:fig1}
\end{center}
\end{figure}

The invariant scattering amplitudes of the processes can be
written as
\begin{eqnarray}
 M_{a}^{\gamma (Z)B_H}&=&\frac{i}{(p_1-k_2)^2}\bar{v}(p_{2})\Lambda_{\mu}^{B_Hee}
\varepsilon^{\mu}(k_{1})(p\!\!/_1-k\!\!/_2)\Lambda_{\nu}^{\gamma
(Z)ee}\varepsilon^{\nu}(k_{2})u(p_{1}),\\ \nonumber
 M_{b}^{\gamma
(Z)B_H}&=&\frac{i}{(p_1-k_1)^2}\bar{v}(p_{2})\Lambda_{\nu}^{\gamma
(Z)ee}\varepsilon^{\nu}(k_2)(p\!\!/_1-k\!\!/_1) \Lambda_{\mu}^{B_H
ee}\varepsilon^{\mu}(k_{1})u(p_{1}).
\end{eqnarray}
The initial electron and positron are denoted by $u(p_{1})$ and
$\bar{v}(p_{2})$, the final states $B_H$ and $\gamma (Z)$ are
presented as $\varepsilon_{\mu}(k_1)$ and
$\varepsilon_{\nu}(k_2)$, respectively.

  \section{The numerical results and conclusions }
 From the scattering amplitudes shown in equation (7), we can see that there are
three new free parameters of the LH model involved in the
scattering amplitudes, i.e., the vev $f$, the mixing parameters
$c'$ and $c$. The custodial $SU(2)$ global symmetry is explicitly
broken, which can generate large contributions to the electroweak
observables. If one adjusts that the SM fermions are charged only
under $U(1)_1$, there exist global severe constraints on the
parameter space of the LH model\cite{constraint}. But if the SM
fermions are charged under $U(1)_{1}\times U(1)_{2}$, the
constraints become relaxed. The scale parameter $f=1-2$ TeV is
allowed for the mixing parameters $c'$ and $c$ in the ranges of
$0.62-0.73$ and $0-0.5$, respectively\cite{Csaki, limit}. To
obtain numerical results of the cross sections, we take into
account the constraints on the parameters of the LH model, and fix
the SM input parameters as $s_{W}^{2}$=0.23, $M_Z=91.187$ GeV,
$v=246$ GeV. The electromagnetic fine-structure constant $\alpha$
at a certain energy scale is calculated from the simple QED
one-loop evolution with the boundary value
$\alpha=\frac{1}{137.04}$\cite{Donoghue}. On the other hand, we
put the kinematic cuts on the final states in the calculation of
the cross sections, i.e., $|y|< 2.5,~p_T> 20$ GeV.

From the equations(5-6), we can see that both the coupling
$B_He^+e^-$ and the $B_H$ mass are strongly depended on the mixing
parameter $c'$, so the cross sections of  $e^+e^- \rightarrow
\gamma(Z)B_H$ should be sensitive to the $c'$. In Fig.2, we plot
the cross sections as the function of $c'(c'=0.62-0.73)$, and take
$f=2$ TeV, $c=0.5$, c.m. energy $\sqrt{s}=800$ GeV as the
examples. It is shown in Fig.2 that the cross sections vanish at $
c'=\sqrt{\frac{2}{5}}$ because the coupling $B_He^+e^-$ becomes
decoupled in this case. In the range $c'>\sqrt{\frac{2}{5}}$, the
cross sections sharply increase with $c'$ increasing. On the other
hand, we find that the cross section of $\gamma B_H$ production is
much large than that of $ZB_H$ production. In wide range of the
parameter space, the cross sections are at the level from tens fb
to one hundred fb for $\gamma B_H$ production and from a few fb to
tens fb for $ZB_H$ production.
\begin{figure}
\begin{center}
\includegraphics [scale=0.7] {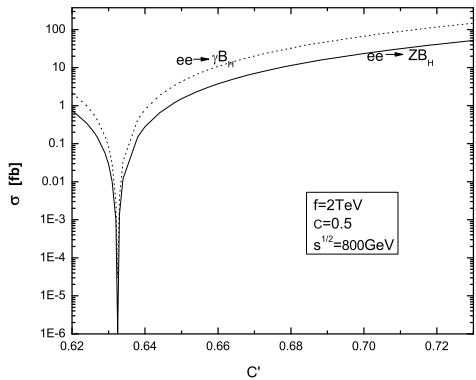}
\caption{The cross sections of the processes $e^+e^- \rightarrow
\gamma (Z)B_H$ as a function of the mixing parameter $c'$, with
$\sqrt{s}=800$ GeV, $c$=0.5, and $f=2$ GeV.}
 \label{fig:fig2}
\end{center}
\end{figure}

The influence of $f$ on the cross sections is also significant.
 Fig.3 shows the plots of cross sections versus $f(f=1-5$ TeV),
with $\sqrt{s}=800$ GeV, $c'=0.68$, and $c=0.5$. With the $f$
increasing, the $B_H$ mass increases and the cross sections
sharply decrease when the $B_H$ mass approaches the kinetic
threshold value.

\begin{figure}
\begin{center}
\includegraphics [scale=0.7] {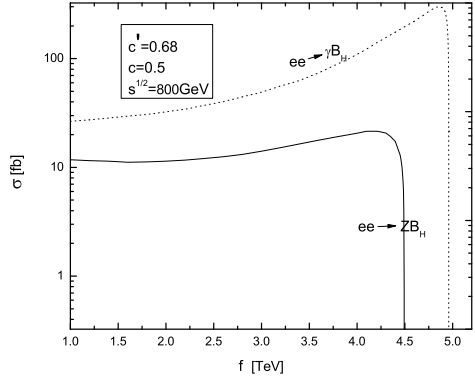}
\caption{The cross sections of the processes $e^+e^- \rightarrow
\gamma (Z)B_H$ as a function of the scale $f$, with $\sqrt{s}=800$
GeV, $c'=0.68$, and $c$=0.5.} \label{fig:fig3}
\end{center}
\end{figure}

The mixing parameter $c$ only has a little effect on the masses of
the final states $B_H$ and Z, so the cross sections are
insensitive to the parameter $c$ and we fix c=0.5 as a example in
our calculation.

In order to give more information about the $\gamma(Z)B_H$
productions, we also plot the angular distributions of these
processes in Fig.4. Where $\theta$ is the angle between the
incoming electron beams and the scattering $B_{H}$. The fig.4
shows that the angular distributions sharply increase when
$cos\theta$ approaches 1 or -1 due to the t-channel resonance
effect. This means that the $B_H$ signals are more concentrated
near to the incoming $e^+e^- $ axis.

\begin{figure}
\begin{center}
\includegraphics [scale=0.7] {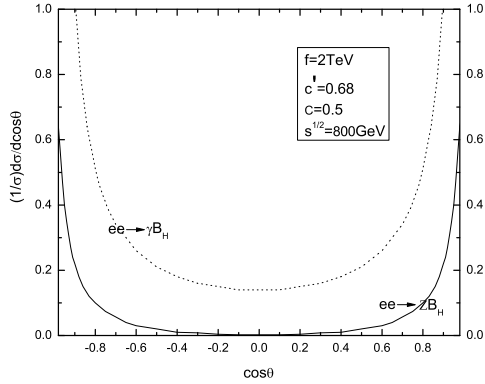}
\caption{The angular distributions of the
   processes $e^+e^- \rightarrow \gamma (Z)B_H$, with
$\sqrt{s}=800$ GeV, $c'=0.68$, $c$=0.5, and $f=2$ GeV. Here
$\theta$ is the angle between the outgoing gauge boson $B_H$ and
the incoming electron.}
 \label{fig:fig4}
\end{center}
\end{figure}

\indent As we have discussed above, with the integrated luminosity
500 $fb^{-1}$ at the ILC, a large number of $B_H$ can be produced
via the processes $e^+e^-\rightarrow \gamma (Z)B_H$ in the wide
range of parameter space of the LH model, specially for the
process $e^+e^-\rightarrow \gamma B_H$. However, the event rate of
$B_H$ identified not only depends on the cross section, but also
depends on the reconstruction efficiencies of the decay channels
of $B_H$. The final states of the $\gamma(Z)B_H$ productions
should include two jets. One is photon jet or the jet decaying
from Z(such jet should includes light quark pair or lepton pair).
Another jet is just the final states decaying from $B_H$. Both
$\gamma$ and $Z$ can be easily identified experimentally, and such
identification is necessary which can depress the SM background
efficiently. To identify $B_H$ from its final states, we also need
to study the decay modes of $B_H$. The main decay modes of $B_H$
are $e^+e^-+\mu^+\mu^-+\tau^+\tau^-, d\bar{d}+s\bar{s},
u\bar{u}+c\bar{c}, ZH, W^+W^-$. The decay branching ratios of
these modes have been studied in reference\cite{Han}. Because the
light lepton pairs $l^+l^-(l=e, \mu)$ are typically well isolated
from all other particles with high efficiency and the number of
$l^+l^-$ background events with such a high invariant mass is very
small, the peak in the invariant mass distribution of $l^+l^-$
should be sensitive to the presence of $B_H$. So the decay modes
$l^+l^-$ are the most ideal modes to detect $B_H$ in most case.
For these leptonic decay modes, the final states of $\gamma B_H$
production should be $\gamma l^+l^-$. In this case, the main SM
background arises from the process $e^+e^-\rightarrow \gamma Z$
with large production rate(at the level of a few pb for
$\sqrt{s}=800$ GeV\cite{Zr}), which can lead to similar multi-jet
topologies. However, it should be very easy to distinguish $B_H$
from $Z$ by measuring the invariant mass distributions of the
$l^+l^-$ because such invariant mass distributions between $B_H$
and $Z$ are significantly different. The measurement of this
lepton pair invariant mass distributions can drastically reduce
the background and the production mode $e^+e^-\rightarrow \gamma
B_H\rightarrow \gamma l^+l^-$ can achieve a very clean SM
background. For the production mode $e^+e^-\rightarrow
ZB_H\rightarrow Zl^+l^-$, the main SM background arises from the
processes $e^+e^-\rightarrow ZZ, ZH$. As we have mentioned above,
$B_H$ can be easily distinguished from $Z$ via their decay modes
$l^+l^-$, and the decay branching ratios of $H\rightarrow l^+l^-$
are very small. Therefore, a clean SM background can also be
achieved if one detect $B_H$ via the production mode
$e^+e^-\rightarrow ZB_H\rightarrow Zl^+l^-$. When the parameter
$c'$ is near $\sqrt{\frac{2}{5}}$, the couplings $B_Hl^+l^-$
become decoupled and the decay modes $l^+l^-$ can not be used to
detect $B_H$. In this case, the decay modes $W^+W^-, ZH$ can
provide a complementary method to probe $B_H$. The decay branching
ratios of $W^+W^-, ZH$ greatly increase when $c'$ is near
$\sqrt{\frac{2}{5}}$, and in this case we might assume enough
$W^+W^-$ and $ZH$ signals to be produced with high luminosity. The
decay mode $Z\rightarrow W^+W^-$ is of course kinematically
forbidden in the SM but $H\rightarrow W^+W^-$ is the dominant
decay mode with Higgs mass above 135 GeV(one or both of W is
off-shell for Higgs mass below 2$M_W$). So the dominant background
for the signal $ZW^+W^-$ arises from the Higgsstrahlung process
$e^+e^-\rightarrow ZH\rightarrow ZW^+W^-$ which is at the order of
tens fb\cite{ZH-background}. Such background would be serious if
one can not distinguish the $W^+W^-$ invariant mass distribution
between H and $B_H$. However, the signal $\gamma W^+W^-$ does not
suffer from such large background problem which would be one
advantage of the process $e^+e^-\rightarrow \gamma B_H$. For
$B_H\rightarrow ZH$, the main final states of $B_H$ are
$l^+l^-b\bar{b}$. Two b-jets reconstruct to the Higgs mass and a
$l^+l^-$ pair reconstructs to the Z mass. On the other hand, the
decay mode $ZH$ involves the off-diagonal coupling $HZB_H$ and the
experimental precision measurement of such off-diagonal coupling
is more easier than that of diagonal coupling. So, the decay mode
$ZH$ would provide an ideal way to verify the crucial feature of
quadratic divergence cancellation in Higgs mass, furthermore such
signals would provide crucial evidence that an observed new gauge
boson is of the type predicted in the little Higgs models. For the
signal $\gamma ZH,~ZZH$, although the same final states can be
produced via $e^+e^-$ collision in the SM, the cross sections of
these processes in the SM are small and the feature that there
exists a peak in the $ZH$ invariant mass distribution for the
signal can further help one to depress such background.

 \indent In summary, with the mass in the range of hundreds GeV,
 the $U(1)$ gauge boson $B_H$ is the lightest one among the new
 gauge bosons in the LH model. Such particle would be accessible in the first running of the ILC and
 provide an earliest signal of the LH model. In this paper, we study the $B_H$ production
 processes associated with a SM gauge boson $Z$ or $\gamma$ via $e^+e^-$ collision, i.e.,
 $e^+e^-\rightarrow \gamma (Z)B_H$. We find that the cross sections are very sensitive to the
 parameters $c',f$ and the cross section of $\gamma B_H$ production is much larger than
 that of $ZB_H$ production. In a wide range of the parameter
space, sufficient events can be produced to detect $B_H$ via these
processes. The signals are more concentrated near to the incoming
$e^+e^- $ axis. In most case, $B_H$ can be detected via its decay
modes $e^+e^-, \mu^+\mu^-$ which can provide the typical signal
and clean background. Therefore, the processes $e^+e^-\rightarrow
\gamma (Z)B_H$ would open an ideal window to probe $B_H$ with the
high luminosity at the planned ILC. Furthermore, if such gauge
boson is observed, the precision measurement is need which could
offer the important insight for the gauge structure of the LH
model and distinguish this model from alternative theories.

\end{document}